\documentclass[]{spie}  
\usepackage{svg}
 
\usepackage{amsmath,amsfonts,amssymb}
\usepackage{graphicx}
\usepackage[colorlinks=true, allcolors=blue]{hyperref}
\usepackage[normalem]{ulem}

\title{Quality control of the CFRP mirror manufacturing process at NPF
}

\author[a,b]{N. Soto}
\author[a,b]{C. Lobos}
\author[b]{P. Mardones}
\author[a,b,d]{A. Bayo}
\author[b,c]{C. Rozas}
\author[a,b]{S. Castillo}
\author[c,d]{G. Hamilton}
\author[b,c]{L. Pedrero}
\author[a,b,e]{S. Zúñiga-Fernández}
\author[a,b]{K. Maucó}
\author[b,c,d]{H. Hakobyan}
\author[b,c,d]{C. Garc\'ia}
\author[b,c]{M. R. Schreiber}
\author[b,c,d]{W. Brooks}

\affil[a]{Instituto  de  F\'isica  y  Astronom\'ia,  Facultad  de  Ciencias,  Universidad de Valpara\'iso}
\affil[b]{N\'ucleo Milenio de Formaci\'on Planetaria - NPF, Valpara\'iso, Chile}
\affil[c]{Universidad T\'ecnica Federico Santa Mar\'ia, Av. España 1680, Valparaíso, Chile}
\affil[d]{Centro Cient\'ifico Tecnol\'ogico de Valpara\'iso - CCTVal, Valparaíso, Chile}
\affil[e]{European Southern Observatory, Santiago, Chile}

\authorinfo{Further author information: (Send correspondence to N. Soto)\\N. Soto: E-mail: nsoto@npf.cl}

\begin{document} 
\maketitle

\begin{abstract}
The surface quality of replicated CFRP mirrors is ideally expected to be as good as the mandrel from which they are manufactured. In practice, a number of factors produce surface imperfections in the final mirrors at different scales. To understand where this errors come from, and develop improvements to the manufacturing process accordingly, a wide range of metrology techniques and quality control methods must be adopted. Mechanical and optical instruments are employed to characterise glass mandrels and CFRP replicas at different spatial frequency ranges. Modal analysis is used to identify large scale aberrations, complemented with a spectral analysis at medium and small scales. It is seen that astigmatism is the dominant aberration in the CFRP replicas. On the medium and small scales, we have observed that fiber print-through and surface roughness can be improved significantly by an extra resin layer over the replica's surface, but still some residual irregularities are present.

\end{abstract}


\section{INTRODUCTION}
\label{sec:intro}  

Production of low-cost segmented primaries for four to eight meter class infrared telescopes is one of the main technology requirements of the future ground-based infrared interferometers such as the Planet Formation Imager (PFI) \cite{monnier2016architecture}. Developments in the manufacturing process of replicated carbon fiber reinforced polymers (CFRP) mirrors have been made by the collaboration between engineers, astronomers and experimental physicist at Núcleo Milenio de Formación Planetaria (NPF) and Centro Científico y Tecnológico de Valapraíso (CCTVal).

A key part of the development of the production process of CFRP mirrors involves the design and implementation of quality control procedures that allow to guide the manufacturing in a cost-and-time effective manner. These procedures have to be designed to quantify the main known quality issues of CFRP mirrors, such as fiber print-through and large shape deformation, which are specific to CFRP-based mirrors. 
Also, the metrology of the convex mandrels from which the replicas are made must be considered. Although optical methods for the measurement of convex optics are well known, their implementation is not cheap and poorly scalable if a slightly larger optical element is not readily available \cite{burge1997measurement}. 

In this work, we present the current status of the quality control process of lightweight mirror production at NPF. A summary of the manufacturing process is first introduced and the main surface errors of replicated CFRP mirrors are then described. The used metrology procedures and some example results are finally presented with a discussion of future improvements.

\section{PROCESS}

The manufacturing process of replicated CFRP optical surfaces consists of five main steps\cite{romeo2006progress}: First a convex mold, known as mandrel, is polished until a spherical high quality surface is achieved. Then ``prepreg'' laminates of CFRP are layered over the mandrel surface with specific orientations in order to achieve a quasi-isotropic layup, that aims to reduce curing-induced deformations. The layup is then cured in an oven or autoclave using vacuum bagging. 

   \begin{figure} [ht]
   \begin{center}
   \begin{tabular}{c} 
   \includegraphics[width=\textwidth]{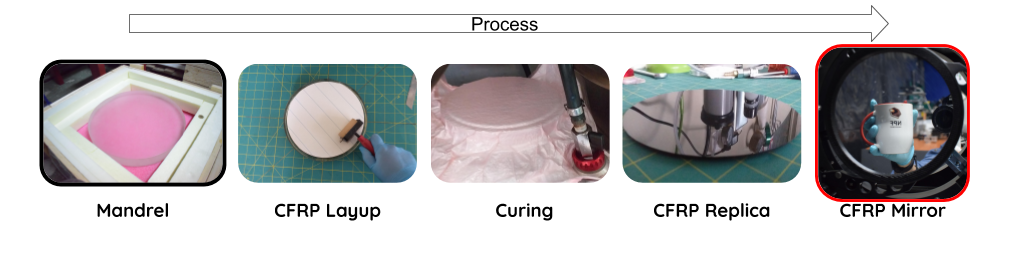}
   \end{tabular}
   \end{center}
   \caption[example] 
   { \label{fig:process} 
The CFRP Mirror production process. Starting from an optically polished convex spherical mandrel, CFRP layers are stacked in a quasi-isotropic layup. Using vacuum bagging the replica is cured. The mirror is obtained after smoothing the surface and a thin layer of metal coating}
   \end{figure} 

As out-of-the-oven replicas present high fiber print-through, it is necessary to apply an extra layer of resin to achieve a smoother surface. The resin is sandwiched between the mandrel and the rough replica to be cured at room temperature to minimise distortions. A thin layer of release agent must be applied over the mandrel to prevent the resin from bonding with it.

   \begin{figure} [ht]
   \begin{center}
   \begin{tabular}{c} 
   \includegraphics[width=.5\textwidth]{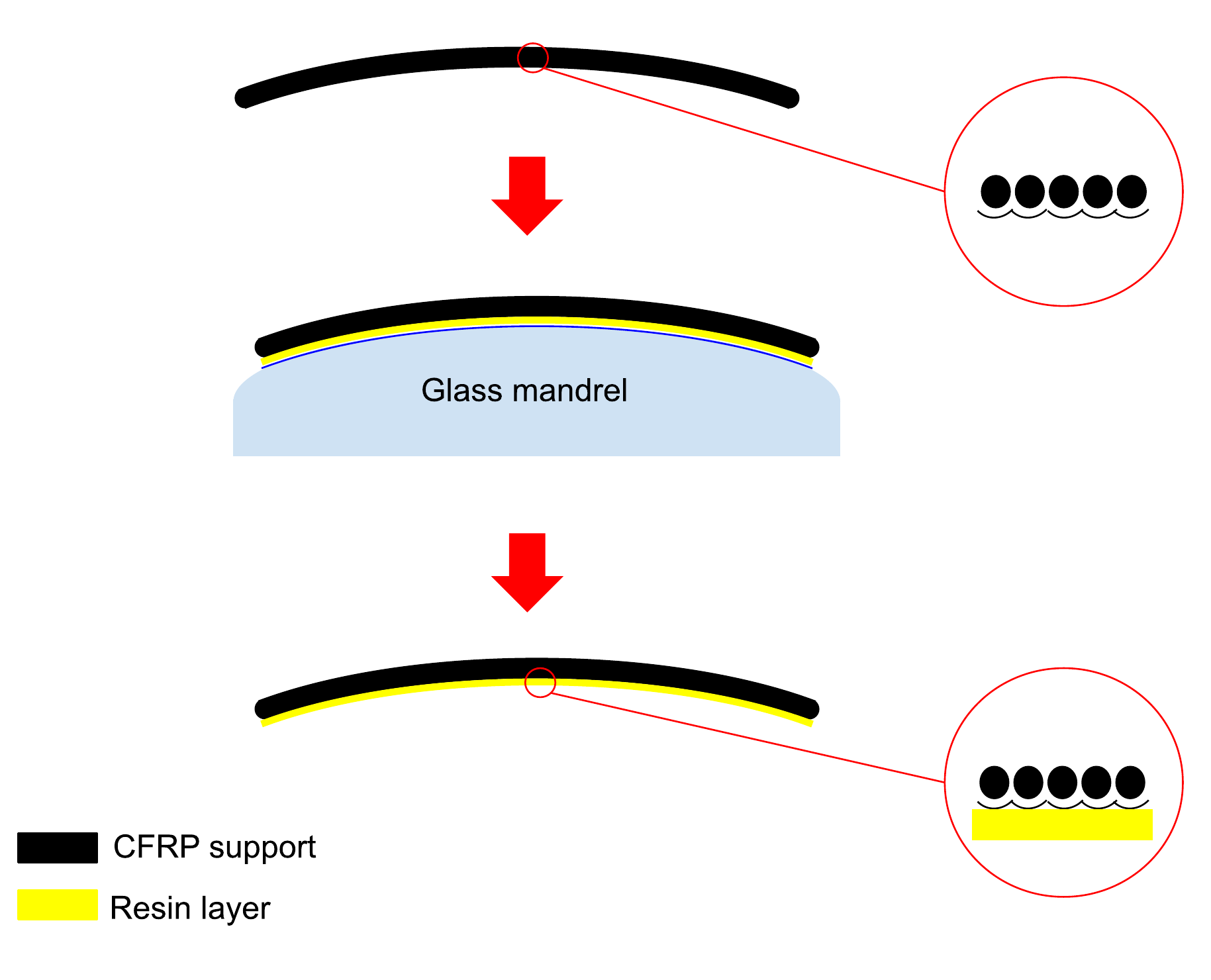}
   \end{tabular}
   \end{center}
   \caption[example] 
   { \label{fig:example} 
Extra resin layer application. The first step shows the rough replica obtained after curing in the oven. Then a liquid resin is sandwiched between the mandrel and the replica. After curing and release from the mandrel, a smooth replica is obtained}
   \end{figure} 

Each step introduces a potential error factor in the replication process that can degrade the optical performance of the final CFRP mirror. Controlling the quality of both mandrel and replicas at every step is then essential to understand the critical factors that induce the predominant errors in the replication process.

\subsection{TYPES AND SOURCES OF SURFACE ERRORS}

During the replication process different factors influence the behaviour of the composite, therefore, the sources of imperfections are varied in nature and ``range of action", adding to a final surface that has artifacts distributed through different scales depending on the underlying physical processes causing them. The main imperfections detectable on a CFRP mirror are: shape distortion, fiber print-through, micro-bubbles and micro-cracking of the polymer matrix.

   \begin{figure} [ht]
   \begin{center}
   \begin{tabular}{c} 
   \includegraphics[width=.6\textwidth]{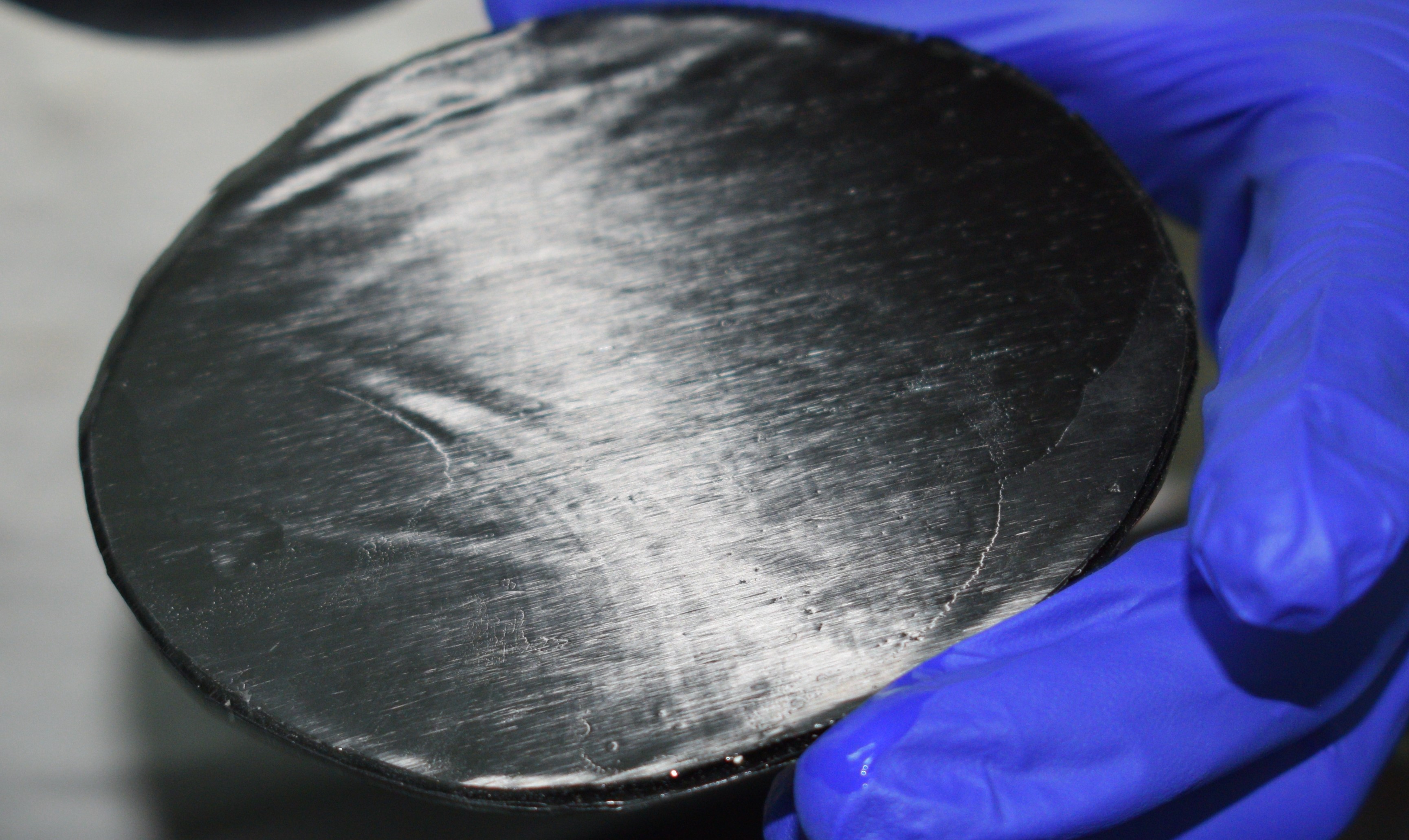}
   \end{tabular}
   \end{center}
   \caption[example] 
   { \label{fig:example} 
One of the first CFRP prototypes showing the typical surface errors: clear fiber print-through, shape distortions and micro-bubbles.}
   \end{figure} 

To identify the causes and quantify the effects of the dominant imperfections it is necessary to measure and analyse the surfaces involved in the process.

The first possible source of imperfections is the mandrel that is being replicated, as it is expected that the final surface quality of the replicas is related directly to that of the mould, at best in a one to one fashion. A careful measurement of the mandrel quality is then needed to assess the expectations of the final result. Nevertheless, optical metrology of convex surfaces is not trivial due to the lack of a real optical focus.  

Mechanical stress during the curing cycle due to layup misalignment and  different thermal behaviour of carbon fibers and epoxy resin is the main source of shape errors in the replicas\cite{thompson2014influence}. Careful alignment of the CFRP layers is necessary to achieve a quasi-isotropic layup that present minimal shape distortion. Even with perfectly aligned layers, fabrication errors of the laminates may induce an-isotropic behaviour. The temperature ramps of the curing process also influence the amount of distortion for a given layup. 

Fibers print-through is a known quality killer of CFRP mirrors \cite{massarello2006fiber}. The structure of the underlying laminates is easily transferred to the surface due to the different thermal behaviour of the polymer matrix and the carbon fibers. Even when applying an extra resin layer to mitigate this effects, residual print-through may still be present in the corrected surface. Its predominance is linked with the thickness of the fibers that are being used, the chemical shrinkage of the epoxy and is also dependent on the laminate type: woven patterns clearly show more print-through when compared to the unidirectional ones. Print-through dominates mid to high spatial frequencies of the surface.

The ``degradation'' level of the polymer, known as matrix state,  can also deteriorate the surface quality of the replicas for high precision replication. Partially cured epoxy will not react as expected in the curing cycle and will produce a range of surface artifacts such as bubbles, voids and micro-crackings. This defects, as far as they are not predominant, will deteriorate the surface in the middle and small scales, increasing the scattering of the reflected light.

Analysing the surface errors by their spatial frequency helps to pinpoint the sources and quantify the effects of the different types of surface imperfections.  Form errors present low spatial frequency deviations from the desired shape. Fiber print-through waviness and roughness  can be seen as clear mid and high frequency surface errors in the direction orthogonal to the fiber's orientation. 

\section{METROLOGY PROCEDURES}

In our lab, metrology procedures have been introduced at different stages of the manufacturing process to quantify the impact of new techniques in the final quality of replicated mirrors. The main parameters to measure are the surfaces shape and their roughness. We are interested in determining the degradation of surface quality from the mandrel to the replicas.

The final optical quality of the mandrel is difficult to quantify due to its convex shape, so we rely on mechanical instruments for its metrology. Sphericity is controlled locally using a ring spherometer. At least 30 points are measured across the surface. Gauge precision of 1 $\mu m$ ensures constant radius of curvature (RoC) within 30 mm accuracy over all the surface for mandrels with RoC = 2400 mm when using a 40 mm ring, as given by the expression:

\begin{equation}
   \Delta  R = -\frac{r^2}{2s^2}\Delta s
\end{equation}

Where $r$ is the radius of the spherometer ring, $s$ the measured sagitta and $\Delta R$ is the uncertainty in the RoC measurement for a given sagitta uncertainty $\Delta s$.  

   \begin{figure} [ht]
   \begin{center}
   \begin{tabular}{c c} 
   \includegraphics[width=.45\textwidth]{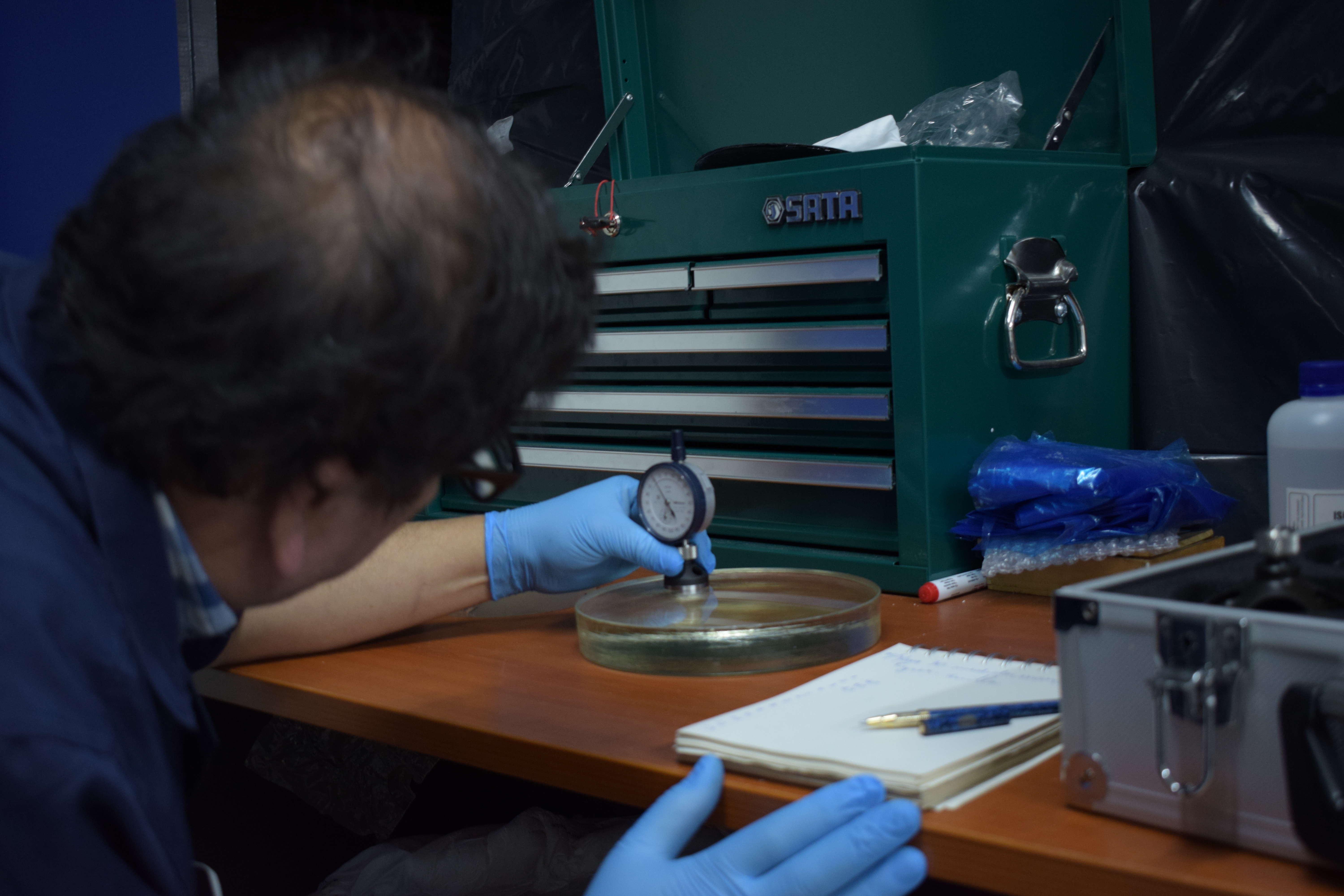} &\includegraphics[width=.43\textwidth]{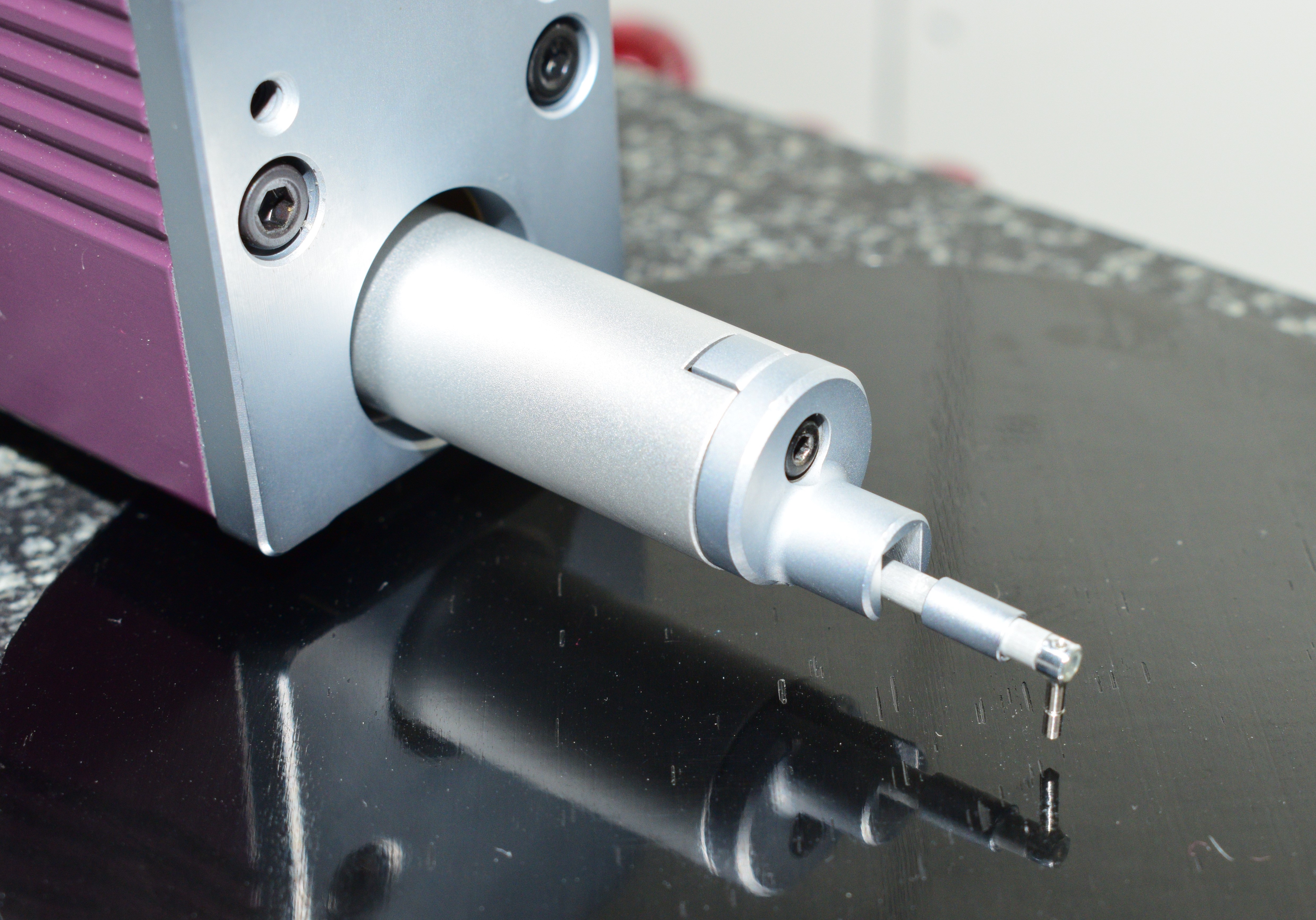}
   \end{tabular}
   \end{center}
   \caption[example] 
   { \label{fig:example} 
Left: Sphericity measurement of the mandrel. If sag measurements are constant within 1 $\mu m$ over the whole surface the mandrel is considered acceptable. Right: CFRP replica surface profiling. Profiles must be perpendicular to the fibers orientation of the fisrt CFRP layer.}
   \end{figure} 

Residual scratches and dents from the polishing process spotted by visual inspection are periodically controlled via microscopy to evaluate surface degradation. Polarised light is also used to evaluate the material stress of transparent mandrels. 

Efforts are being made to develop a custom optical characterisation method for large convex spheres. In the meantime, due to the homogeneity of the glass mandrel surface produced by traditional polishing, mechanical measurements seem sufficient to provide a characterisation of its quality. Nevertheless, in a production environment, quicker methods will be necessary.

Improving the surface smoothness is a key challenge in the development of CFRP mirrors, as surface irregularities have a negative impact in the performance of optical elements due to the scattering of the incident light. A good parameter to establish the surface quality is its roughness, as it determines the ratio of specular to total reflected radiation by the relation\cite{bennett1961relation}:

\begin{equation}
 \frac{R_s}{R_t} = e^{-(4\pi\frac{\sigma}{\lambda})^2}    
\end{equation}

Where $R_s$ is the specular reflectance, $R_t$ is the total reflectance of a given material, $\sigma$ is the surface roughness and $\lambda$ is the radiation wavelength. For example, to achieve a 95 \% of specular reflection in the visible range ($\lambda = 532 $ nm), it is necessary to achieve a surface roughness $\sigma < 10 $ nm. 

It is important to recall that surface irregularities are distributed at different scales. As mentioned in the previous chapter, CFRP replicas are dominated by the wavy pattern of the fibers rather than random homogeneous surface roughness. Measurements of surface quality have to consider scales in the order of the material's patterns dimensions\cite{aikens2008specification}, i.e. in the 0.01 - 10 mm range for CFRP to cover features arising from single fibers to packs of a few thousands of them (which is the typical manufacturing tow size). We therefore divide our measurements in short and mid spatial frequencies to distinguish between the random roughness of the polymer and the wavy pattern of the fibers.

Both mid and short spatial frequencies are measured using a mechanical stylus profilometer, the Mitutoyo SJ-410, first during the mandrel fabrication and then for each replica before and after applying the extra resin layer. Environmental vibrations make precise mechanical measurements difficult, so the elements must be placed over a vibration isolation support. We use an optical table with passive vibration dampeners. Twelve 0.8 mm ``short'' profiles are measured at random over the surface to measure its roughness in the 0.002-0.08 mm spatial period range. Four additional 40 mm ``long'' profiles are taken to characterise the waviness in the 0.08-8 mm range. The mandrel is considered acceptable when its roughness is less than 10 nm RMS, as its quality determines the base level to assess the improvement margin for the CFRP replicas.

\begin{figure} [ht]
   \begin{center}
   \begin{tabular}{c} 
   \includegraphics[width=.54\textwidth]{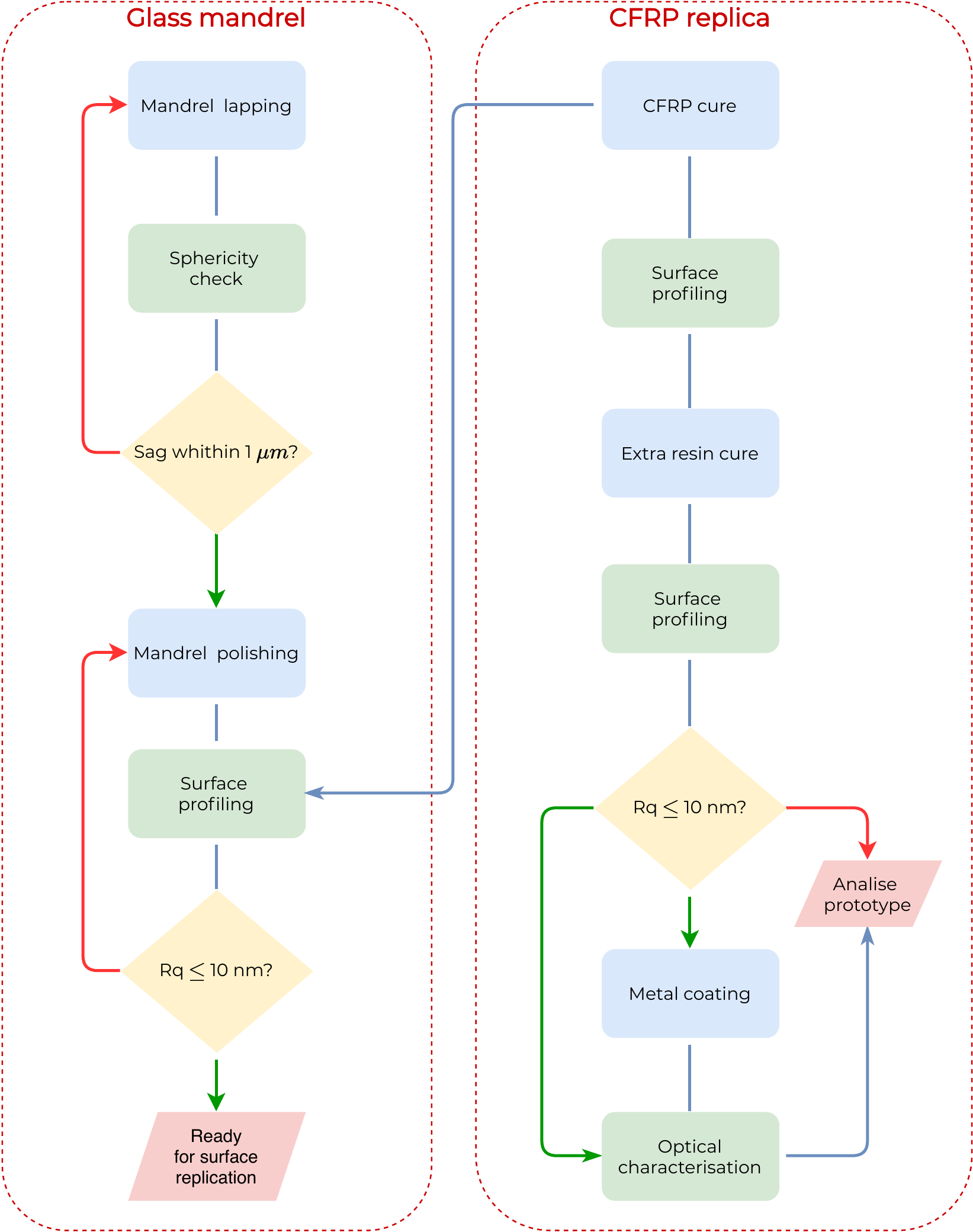}
   \end{tabular}
   \end{center}
   \caption[mandrel] 
   { \label{fig:metrology_process} 
Manufacturing process for the development of CFRP mirrors including metrology procedures.}
   \end{figure}

Replicas are also characterised optically if their quality is sufficient. As the epoxy resin has $\sim$ 5 \% reflectivity, it is posible to perform optical tests without the necessity of a metal coating. Optical characterisation is performed using a Shack-Hartmann wavefront sensor, the Optino-mu, along with other traditional qualitative methods such as Ronchi and Foucault tests. Modal and spectral analysis of the data is finally used to compare the mandrel and replicas surfaces at different scales. Figure \ref{fig:metrology_process} summarises the manufacturing process currently in use in the development of CFRP mirrors at NPF, including the metrology procedures.

\section{RESULTS AND DISCUSSION}

\begin{figure} [ht]
   \begin{center}
   \begin{tabular}{c c} 
   \includegraphics[width=.5\textwidth]{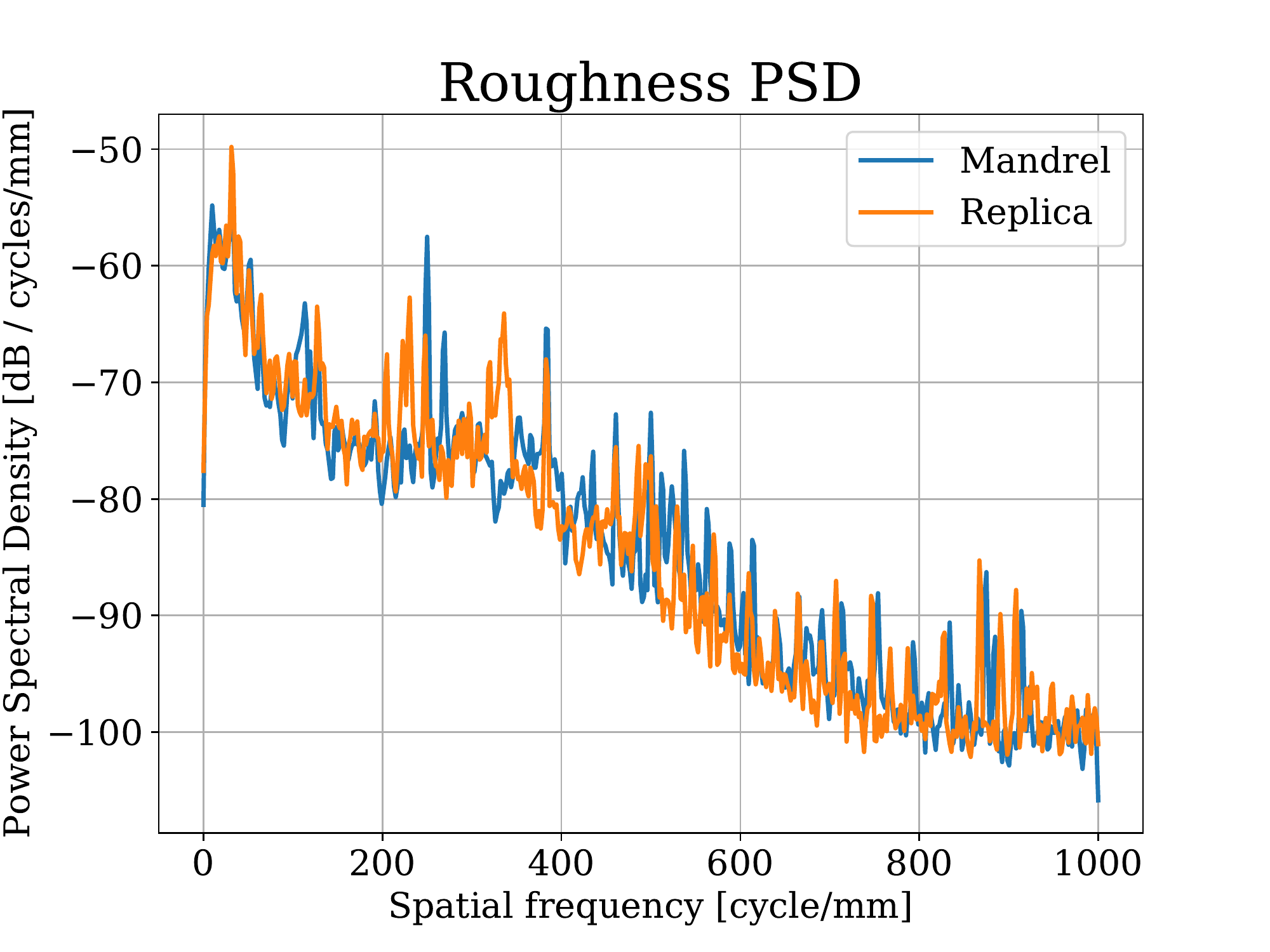} &
   \includegraphics[width=.5\textwidth]{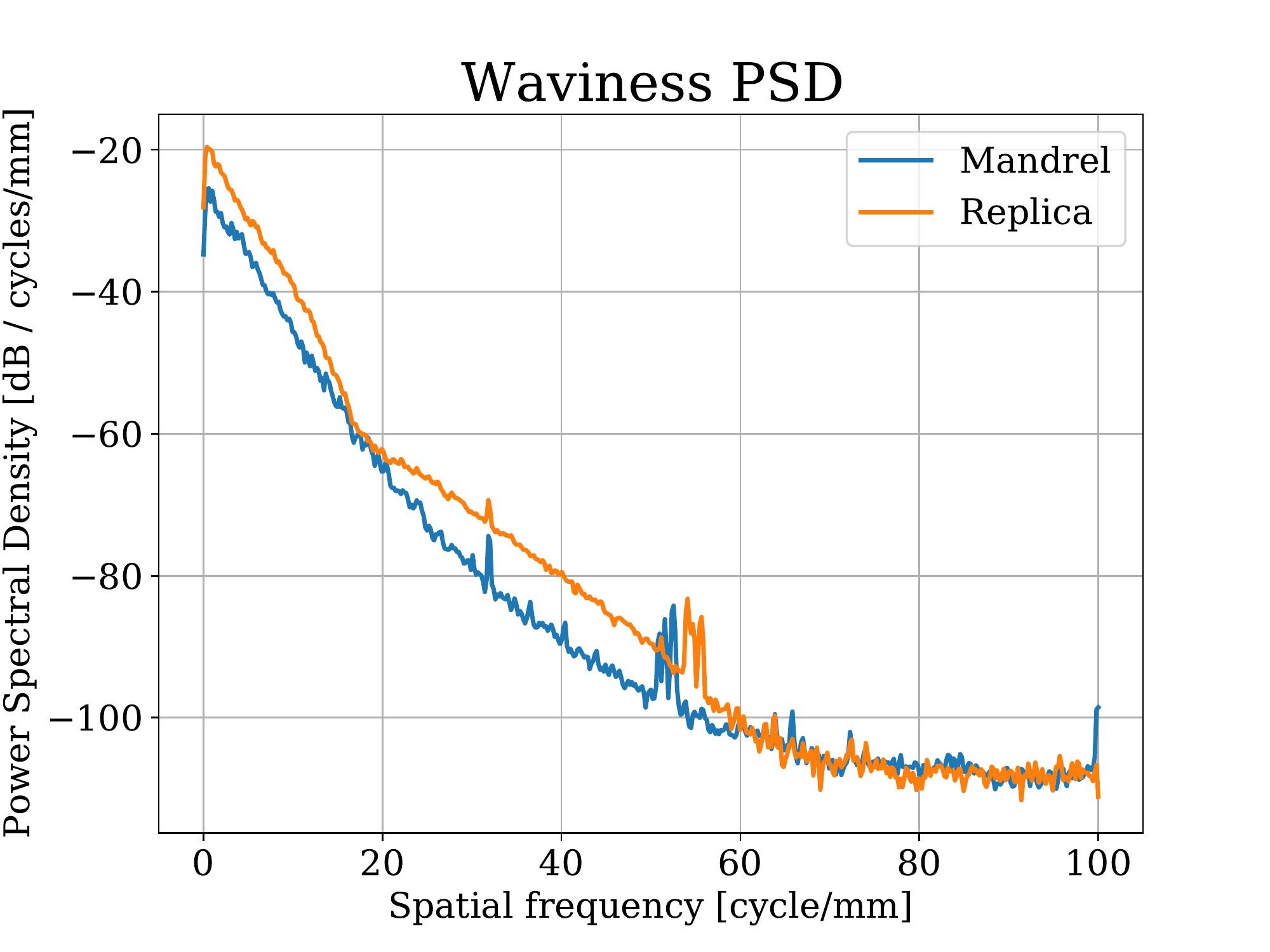}
   \end{tabular}
   \end{center}
   \caption[mandrel] 
   { \label{fig:psds} 
Comparison of mandrel and replica spatial frequencies. High spatial frequency distribution is unaltered in the replication process as shown in the left panel. Mid spatial frequency content of the CFRP replicas is still significant compared with the mandrel as shown in the right panel}
   \end{figure}

So far our tests have shown that, although the application of an extra resin layer improves considerably the CFRP surface, residual irregularities are still an issue in the production of replicated mirrors. Figure \ref{fig:psds} shows the comparison of the power spectral densities for high and mid spatial frequencies measurements of the mandrel and a CFRP replica after the application of the extra resin layer.

It can be seen that  the high spatial frequency content of the roughness profiles is almost equivalent, showing that the cured epoxy resin can achieve very smooth surfaces and is limited only by the mandrel quality. Meanwhile, the CFRP replica shows a larger mid spatial frequency content up to 60 cycles/mm, which is equivalent to larger irregularities with periods down to 0.016 mm, i.e, the width of two fibers.

Given the thickness of the additional epoxy layer ($\sim$ 50 $\mu m$) it is possible to polish the solid resin to improve its surface quality, at the expense of adding steps to the manufacturing process. A new layer orientation layup is also being investigated to smooth mid spatial frequency components from fiber print-through (see Castillo et al. in this same SPIE volume), and a different room temperature curing resin has shown promising results increasing the replication fidelity.

\begin{figure} [ht]
   \begin{center}
   \begin{tabular}{c c} 
   \includegraphics[width=.5\textwidth]{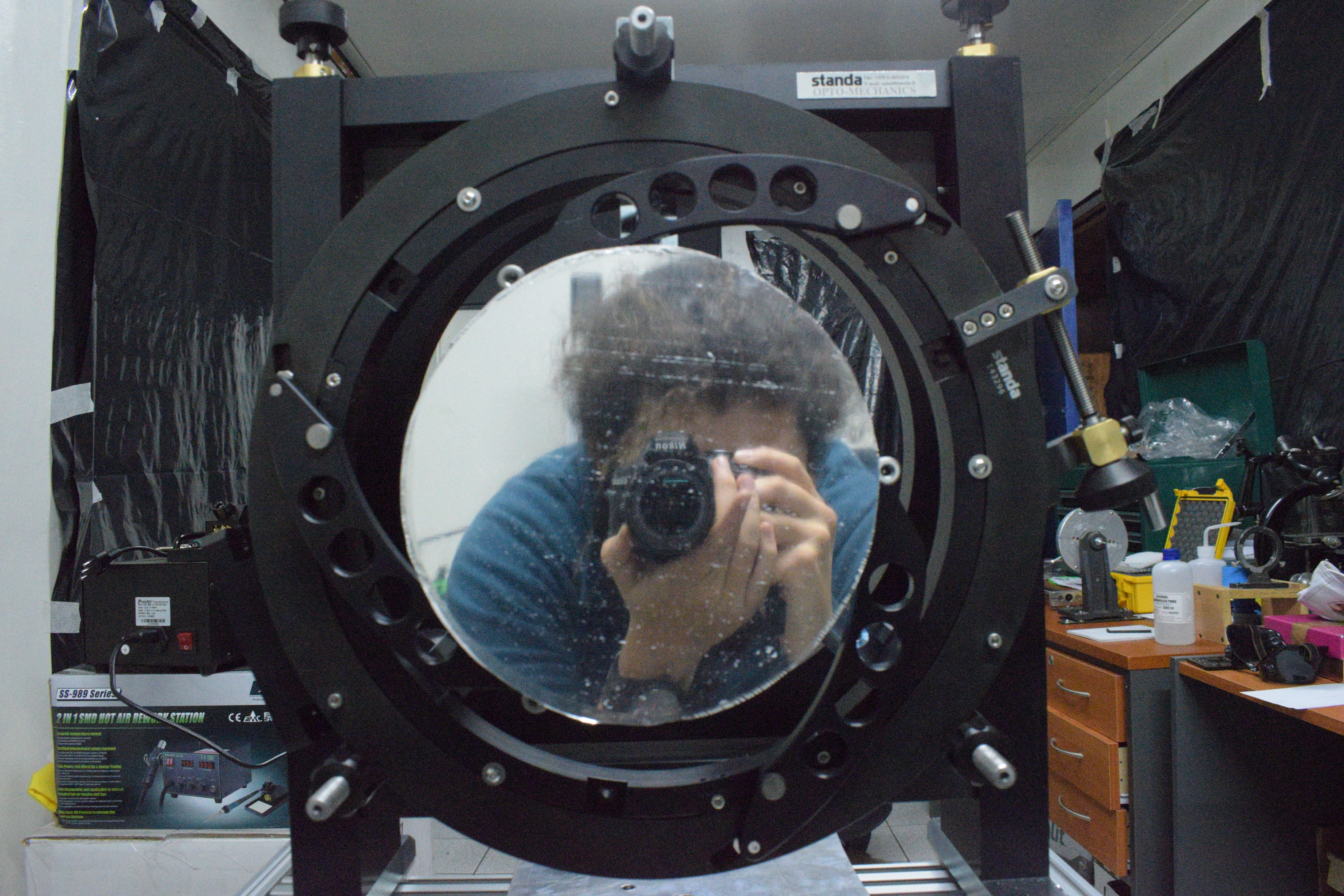} &
   \includegraphics[width=.4\textwidth]{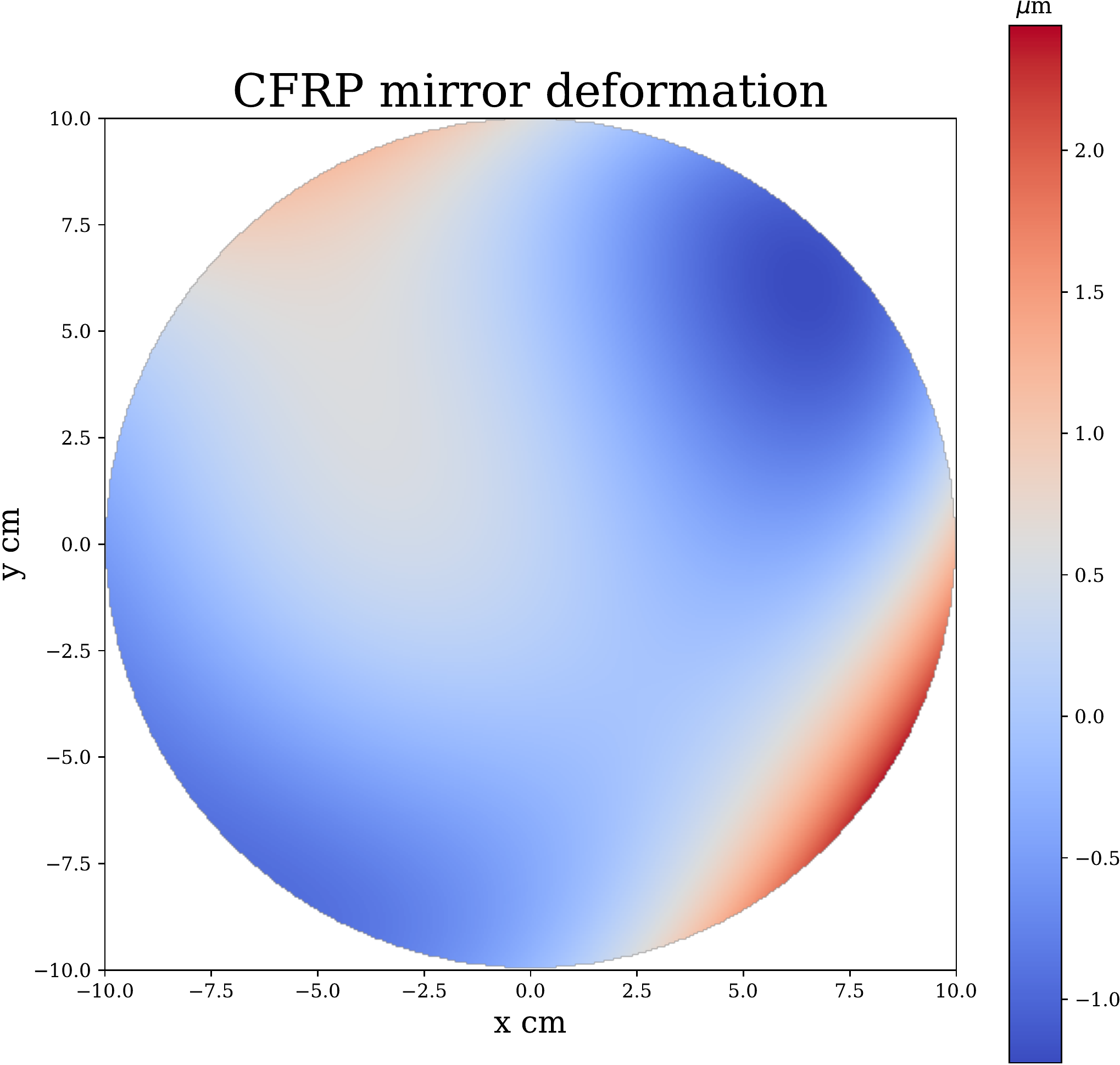}
   \end{tabular}
   \end{center}
   \caption[mandrel] 
   { \label{fig:optical} 
Coated mirror modal surface reconstruction from Shack-Hartmann wavefront sensor: tip, tilt and defocus removed. Astigmatism is the predominant aberration}
   \end{figure}

Shape deformations are also difficult to eliminate given the sensitivity of the process to many variables that influence the behaviour of the CFRP layups. Astigmatism is the principal optical aberration for 20cm CFRP mirrors prototypes, as shown in figure \ref{fig:optical}, caused by the mechanical stress in the curing process. A better control of the temperature ramps and the layer’s alignment is then of critical importance to improve the final mirror surface quality. Nevertheless, due to the high flexibility of the thin CFRP replicas, it is possible to correct curing-induced deformations using adjustable supporting cells\cite{steeves2013ultra}. So far, 3D-printed prototype cells have been proven useful to reduce large scale aberrations (see, Bayo et al. in this SPIE volume)

Another aspect to consider is the degradation over time due to environmental causes\cite{zaldivar2017hygroscopic}. Systematic assessment of mirror degradation in working conditions is critical to establish the potential use of CFRP mirrors in future astronomical telescopes. We have transported a 20cm replica to Paranal observatory and left it exposed to the elements to evaluate its quality degradation over a long period of time. The exposed replica at Paranal was not significantly affected by environmental factors in a one month period, and it will be evaluated periodically to account for long term effects such as  extreme humidity variations. 

\section{CONCLUSIONS}

We have introduced a detailed description of our current metrology procedures for the development of replicated CFRP mirrors. Both mandrel and replicas are measured systematically across the different stages of the manufacturing process. We rely on mostly manual measurements for the development process using well established metrology instruments. The mandrel shape is controlled with an spherometer to ensure a constant sag across its surface. Stylus profilometry is used to characterise the surface quality of both mandrel and mirrors. Final mirrors are measured using traditional optical methods.

So far, we have been able to improve the quality of the mirror prototypes considerably, based on a systematic method of research and process development. Still, further improvements are necessary to meet the requirements for telescope mirrors in the visible to mid infrared range. Also, automation of the different sub-processes, including the metrology, is crucial to reduce the time and cost of producing CFRP mirrors. Implementation of automated procedures, based in the experience gained so far, will be the next challenge in the road to establish a scalable production process.

\acknowledgments 
 
All the authors acknowledge financial support from Iniciativa Cient\'ifica Milenio v\'ia N\'ucleo Milenio de Formaci\'on Planetaria. N.S and G.H acknowledge financial support from the PIIC program at USM. A.B acknowledges support from FONDECYT grant 1190748, A.B. and N.S. acknowledges support from ESO Comit\'e-Mixto and A.B. from QUIMAL funding agencies. M.S., S.C and C.L acknowledge support from the ALMA-CONICYT fund.  S.Z-F acknowledges financial support from the European Southern Observatory via its studentship program and ANID via PFCHA/Doctorado Nacional/2018-21181044. K.M. acknowledges financial support from CONICYT-FONDECYT project no. 3190859

\bibliography{report} 
\bibliographystyle{spiebib} 

\end{document}